\documentstyle[prl,twocolumn,aps,floats,epsf]{revtex}
\newtheorem{theorem}{Theorem}
\newtheorem{definition}[theorem]{Definition}
\newtheorem{prop}[theorem]{Proposition}
\newtheorem{cor}[theorem]{Corollary}

\begin{document}
% \draft

%\twocolumn[\hsize\textwidth\columnwidth\hsize\csname =
%@twocolumnfalse\endcsname

\title{A quantum analog of Huffman coding}

\author{Samuel L.~Braunstein,
Christopher A.\ Fuchs,\\ Daniel Gottesman, and Hoi-Kwong
Lo\thanks{Correspondence should be addressed to H.-K. Lo, e-mail:
hkl@hplb.hpl.hp.com or S.\ L.\ Braunstein, e-mail:
schmuel@sees.bangor.ac.uk.}
\thanks{S.~L. Braunstein is at SEECS, University of Wales, Bangor LL57
1UT, UK,
and at Hewlett-Packard Labs, Filton Road, Stoke Gifford, Bristol
BS34 8QZ, UK}
\thanks{C.~A. Fuchs is at Norman Bridge Laboratory of Physics 12-33,
California Institute of Technology, Pasadena, CA 91125, USA}
\thanks{D.~Gottesman is at T-6, Los Alamos National Laboratory,
Los Alamos, NM 87545, USA}
\thanks{H.-K.~Lo is at Hewlett-Packard Labs, Filton Road, Stoke
Gifford, Bristol BS34 8QZ, UK}
}

\date{\today}
\maketitle

\begin{abstract}
We analyze a generalization of Huffman coding to the quantum case. In
particular, we notice various difficulties in using instantaneous
codes for quantum communication. Nevertheless, for the {\it
storage\/} of quantum information, we have succeeded in constructing
a Huffman-coding inspired quantum scheme. The number of computational
steps in the encoding and decoding processes of $N$ quantum signals
can be made to be of polylogarithmic depth by a massively parallel
implementation of a quantum gate array. This is to be compared with
the $O (N^3)$ computational steps required in the sequential
implementation by Cleve and DiVincenzo of the well-known quantum
noiseless block coding scheme of Schumacher. We also show that
$O(N^2(\log N)^a)$ computational steps are needed for the {\it
communication\/} of quantum information using another Huffman-coding
inspired scheme where the sender must disentangle her encoding device
before the receiver can perform any measurements on his signals.
\end{abstract}

%\begin{keywords}
%Quantum information, data compression, Huffman coding, quantum
%coding, instantaneous codes, variable length codes
%\end{keywords}
%]

\section{Introduction}

There has been much recent interest in the subject of quantum
information processing.  Quantum information is a natural
generalization of classical information.  It is based on quantum
mechanics, a well-tested scientific theory in real experiments.  This
paper concerns quantum information.

%This paper is an attempt
The goal of this paper is to find a quantum source coding scheme
analogous to Huffman coding in the classical source coding theory
\cite{Cover}.  Let us recapitulate the result of classical theory.
Consider the simple example of a memoryless source that emits a
sequence of independent, identically distributed signals each of which
is chosen from a list $w_1, w_2, \cdots ,w_n$ with probabilities $p_1,
p_2, \cdots, p_n$. The task of source coding is to store such signals
with a minimal amount of resources.  In classical information theory,
resources are measured in bits.  A standard coding scheme to use is
the optimally efficient Huffman coding algorithm, which is a
well-known lossless coding scheme for data compression.

Apart from being highly
efficient, it has the advantage of
being instantaneous, i.e., unlike block coding
schemes,
the encoding and decoding
of each signal can be done immediately. Note also that codewords
of variable lengths are used to achieve efficiency.
As we will see below, these two
features---instantaneousness and variable length---of Huffman
coding are difficult to generalize
to the quantum case.

Now let us consider quantum information.  In the {\it quantum\/} case,
we are given a quantum source which emits a time sequence of
independent identically distributed pure-state quantum signals each of
which is chosen from $| u_1 \rangle , | u_2 \rangle, \cdots | u_m
\rangle$ with probabilities $q_1, q_2 \cdots, q_m$, respectively.
Notice that $| u_i \rangle $'s are normalized (i.e., unit vectors) but
not necessarily orthogonal to each other. Classical coding theory can
be regarded as a special case when the signals $| u_i \rangle$ are
orthogonal.  The goal of quantum source coding is to minimize the
number of dimensions of the Hilbert space needed for almost lossless
encoding of quantum signals, while maintaining a high fidelity between
input and output.  For a pure input state $| u_i\rangle $, the
fidelity of the output density matrix $\rho_i$ is defined as the
probability for it to pass a yes/no test of being the state $|
u_i\rangle $. Mathematically, it is given by $\langle u_i | \rho_i |
u_i\rangle $ \cite{fidelity}.  In particular we will be concerned with
the average fidelity $F=\sum_i q_i \langle u_i | \rho_i | u_i\rangle $
It is convenient to measure the dimensionality of a Hilbert space in
terms of the number of qubits (i.e., quantum bits) composing it; that
is, the base-2 logarithm of the dimension.

Though there has been some preliminary work on quantum Huffman coding
\cite{schpri}, the most well-known quantum source coding scheme is a
block coding scheme \cite{schumacher,jozsa}.  The converse of this
coding theorem was proven rigorously in \cite{fuchs}.  In block
coding, if the signals are drawn from an ensemble with density matrix
$\rho =\sum q_j |u_j\rangle \langle u_j |$, Schumacher coding,
which is almost lossless,
compresses $N$ signals into $N S(\rho)$ qubits, where $S(\rho) = -
{\rm tr}\ \rho \log \rho$ is the von Neumann entropy.  To encode $N$
signals {\it sequentially}, it requires $O(N^3)$ computational steps
\cite{cleve}. The encoding and decoding processes are far from
instantaneous. Moreover, the lengths of all the codewords are the
same.

\section{Difficulties In A Quantum Generalization}
A notable feature of quantum information is that measurement of it
generally leads to disturbance.  While measurement is a passive
procedure in classical information theory, it is an integral
part of the formalism of quantum mechanics and is an active process.
Therefore, a big challenge in quantum coding is: How to encode and
decode without disturbing the signals too much by the measurements
involved?
To illustrate the difficulties involved,
we shall first attempt a naive generalization of Huffman coding to
the quantum case.
Consider the density matrix for each signal $\rho =\sum q_j
| u_j \rangle \langle u_j |$ and diagonalize it into
\begin{equation}
\rho=\sum_i p_i | \phi_i \rangle \langle \phi_i | \;,
\end{equation}
where $| \phi_i \rangle$ is an eigenstate and the eigenvalues $ p_i$'s
are arranged in decreasing order.  Huffman coding of a corresponding
classical source with the same probability distribution $p_i$'s allows
one to construct a one-to-one correspondence between Huffman codewords
$h_i $ and the eigenstates $ | \phi_i \rangle $.  Any input quantum
state $| u_j\rangle$ may now be written as a sum over the complete set
$|\phi_i\rangle$. Remarkably, this means that, for such a naive
generalization of Huffman coding, the length of each signal is a
quantum mechanical variable with its value in a superposition of the
length eigenstates.  It is not clear what this really means nor how to
deal with such an object. If one performs a measurement on the length
variable, the statement that measurements lead to disturbance means
that irreversible changes to the $N$ signals will be introduced which
disastrously reduce the fidelity.

Therefore, to encode the signals faithfully, the sender and the
receiver are forbidden to measure the length of each signal.  We
emphasize that this difficulty---that the sender is ignorant of the
length of the signals to be sent---is, in fact, very general.  It
appears in any distributed scheme of quantum computation. It is also
highly analogous to the synchronization problem in the execution of
subroutines in a quantum computer: A quantum computer program runs
various computational paths simultaneously. Different computational
paths may take different numbers of computational steps. A quantum
computer is, therefore, generally unsure whether a subroutine has
been completed or not. We do not have a satisfactory resolution to
those subtle issues in the general case.  Of course, the sender can
always avoid this problem by adding redundancies (i.e., adding enough
zeroes to the codewords to make them into a fixed length).  However,
such a prescription is highly inefficient and is self-defeating for
our purpose of efficient quantum coding. For this reason, we reject
such a prescription in our current discussion.

In the hope of saving resources, the natural next step to try is to
stack the signals in line in a single tape during the transmission.
To greatly simplify our discussion we shall suppose that the
read/write head of the machine is quantum mechanical with its location
given
by an internal state of the machine (this head location could be thought
of as being specified on a separate tape).
But then the second problem arises.
Assuming a fixed speed of transmission, the receiver can never be sure
when a particular signal, say the seventh signal, arrives. This is
because the {\it total\/} length of the signals up to that point
(from the first to seventh signals) is a quantum mechanical variable
(i.e., it is in a superposition of many possible values).
Therefore, Bob generally has a hard time in deciding when
would be the correct instant to decode the
seventh signal in an instantaneous quantum code.

Let us suppose that the above problem can be solved.
For example, Bob may wait ``long enough'' before performing any
measurements. We argue that
there remains a third difficulty which
is fatal for {\it instantaneous\/} quantum
codes---that the head location of the encoder is {\it entangled\/} with
the total length of the signals.
If the decoder consumes the quantum signal (i.e.,
performs measurements on the signals) before the encoding is completed,
the record of the total length of the signals
in the encoder head will destroy quantum coherence.
This decoherence effect is physically the same as
a ``which path'' measurement that
destroys the interference pattern in a double-slit experiment.
One can also understand this effect
simply by considering an example of
$N$ copies of a state $a | 0 \rangle + b | 1 \rangle$.
It is easy to show that if the encoder couples an encoder head to
the system and keeps a record of
the total number of zeroes, the state of each signal will become
impure. Consequently, the fidelity between the input and the output is
rather poor.

\section{Storage of Quantum Signals}

Nevertheless, we will show here that Huffman-coding inspired quantum
schemes do exist for both storage and communication of quantum
information.  In this section we consider the problem of storage.
Notice that the above difficulties are due to the requirement of
instantaneousness.  This leads in a natural way to the question of
{\it storage\/} of quantum information, where there is no need for
instantaneous decoding in the first place.  In this case, the decoding
does not start until the whole encoding process is done. This
immediately gets rid of the second (namely, when to decode) and third
(namely, the record in the encoder head) problems mentioned in the
last section.  However, the first problem reappears in a new
incarnation: The {\it total\/} length of say $N$ signals is unknown
and the encoder is not sure about the number of qubits that he should
use.  A solution to this problem is to use essentially the law of
large numbers.  If $N$ is large, then asymptotically the length
variable of the $N$ signals has a probability {\it amplitude\/}
concentrated in the subspace of values between $N ( \bar{L}- \delta)$
and $N ( \bar{L} + \delta)$ for any $\delta > 0$
\cite{schumacher,jozsa,fuchs}.  Here $\bar{L}$ is the weighted average
length of a Huffman codeword.  One can, therefore, truncate the signal
tape into one with a {\it fixed\/} length say $N (\bar{L} +
\delta)$. [`0's can be padded to the end of the tape to make up the
number if necessary.]  Of course, the whole tape is not of variable
length anymore.  Nonetheless, we will now demonstrate that this tape
can be a useful component of a new coding scheme---which we shall call
quantum Huffman coding---that shares some of the advantages of Huffman
coding over block coding. In particular, assuming that quantum gates
can be applied in {\it parallel}, the encoding and decoding of quantum
Huffman coding can be done efficiently.  While a sequential
implementation of quantum source {\it block\/} coding
\cite{schumacher,jozsa,fuchs} for $N$ signals requires $O (N^3)$
computational steps \cite{cleve}, a parallel implementation of quantum
Huffman coding has only $O ( ({\log N})^a)$ depth for some positive
integer $a$.

We will now describe our coding scheme
for the storage of quantum signals.
%As before, we are given
%a quantum source which emits a time
%sequence of independent identically distributed quantum signals each of
%which is $| u_1 \rangle, | u_2 \rangle, \cdots,| u_n \rangle$ with
%probabilities $q_1, q_2 \cdots, q_n$ respectively.
%We construct the density matrix for each signal $\rho =\sum q_j
%| u_j \rangle \langle u_j |$ and diagonalize it into
%\begin{equation}
%\rho=3D \sum_{i=3D1}^n
% p_i | \phi_i \rangle \langle \phi_i | \;,
%\end{equation}
%where $p_i$ is an eigenvalue and $| \phi_i \rangle$ an eigenstate.
As before, we consider a quantum source emitting a sequence of
independent identically distributed quantum signals with a density
matrix for each signal shown in Eq.\ (1) where $p_i$'s are the
eigenvalues.  Considering Huffman coding for a classical source with
probabilities $p_i$'s allows one to construct a one-to-one
correspondence between Huffman codewords $h_i $ and the eigenstates $|
\phi_i \rangle $.  For parallel implementation, we find it useful to
represent $| \phi_i \rangle$ by two pieces,\footnote{The second piece
contains no new information. However, it is useful for a massively
parallel implementation of the shifting operations, which is an
important component in our construction.}  the first being the Huffman
codeword, padded by the appropriate number of zeroes to make it into
constant length,\footnote{The encoding process to be discussed below
will allow us to reduce the total length needed for $N$ signals.}  $|
0 \cdots 0 h_i\rangle$, the second being the length of the Huffman
codeword, $| l_i \rangle$, where $l_i = {\rm length} ( h_i) $. We also
pad zeroes to the second piece so that it becomes of fixed length
$\lceil \log l_{\rm max} \rceil$ where $l_{\rm max}$ is the length of
the longest Huffman codeword.  Therefore, $| \phi_i \rangle $ is
mapped into $| 0 \cdots 0 h_i\rangle | l_i \rangle$. Notice that the
length of the second tape is $ \lceil \log l_{\rm max} \rceil $ which
is generally small compared to $n$.  The usage of the second tape is a
small price to pay for efficient parallel implementation.

In this Section, we use the model of a quantum gate array for quantum
computation.  The complexity class {\bf QNC} is the class of quantum
computations that can be performed in polylogarithmic parallel
depth~\cite{moore}.  We prove the following theorem:

\begin{theorem}
Encoding or decoding of a quantum Huffman code for storage is in the
complexity class {\bf QNC}.
\end{theorem}

The proof follows in the next two subsections.

\subsection{Encoding}

Without much loss of generality, we suppose that the total number of
messages is $N=2^r$ for some positive integer $r$.  We propose to
encode by divide and conquer.  Firstly, we divide the messages into
pairs and apply a merging procedure to be discussed in
Eq.~(\ref{merging}) to each pair.  The merging effectively reduces the
total number of messages to $2^{r-1}$.  We can repeat this process.
Therefore, after $r$ applications of the merging procedure below, we
obtain a single tape containing all the messages (in addition to the
various length tapes containing the length information).

The first step is the merging of two signals into a single message.
Let us introduce a message tape.  For simplicity, we simply denote $|
0 \cdots 0 h_{i_1}\rangle $ by $| h_1\rangle$, etc.

\begin{eqnarray}
   ~& | h_1\rangle | l_1 \rangle |h_2\rangle | l_2 \rangle &
      | {\bf 0} \rangle_{\rm tape} \nonumber \\
\stackrel{\rm swap}{\longrightarrow} & | {\bf 0} \rangle | l_1 \rangle
      | h_2\rangle | l_2 \rangle &
      | 0 \cdots 0 h_1 \rangle_{\rm tape} \nonumber \\
\stackrel{\rm shift}{\longrightarrow} &  | {\bf 0} \rangle| l_1 \rangle
      | h_2\rangle | l_2 \rangle &
      | h_1  0 \cdots 0\rangle_{\rm tape} \nonumber \\
\stackrel{\rm swap}{\longrightarrow} & | {\bf 0} \rangle | l_1 \rangle
      | {\bf 0} \rangle | l_2 \rangle &
      | h_1  0 \cdots 0 h_2\rangle_{\rm tape} \nonumber \\
\stackrel{\rm shift}{\longrightarrow} & | {\bf 0} \rangle | l_1 \rangle
      | {\bf 0} \rangle | l_2 \rangle &
      | h_1 h_2 0 \cdots 0 \rangle_{\rm tape}  \;.
\label{merging}
\end{eqnarray}

We remark that the swap operation between any two qubits can be done
efficiently by using an array of three XOR's with the two qubits
alternately used as the control and the target.\footnote{In
equation~(\ref{merging}), we do not include the position of the head,
since it is simply dependent on the sum of the message lengths and can
be reset to 0 after the process is completed.}  The shift operation is
just a permutation and therefore can be done in constant
depth~\cite{moore}.  However, we actually need something slightly
stronger: a controlled-shift, controlled by functions of the lengths
$|l_1\rangle$ and $|l_2\rangle$, which are quantum variables.  To do a
shift controlled by the register $|s\rangle$, we expand $s$ in binary,
and perform a shift by $2^i$ positions conditioned on the appropriate
bit of $s$.  When $|s\rangle$ is a quantum register in a
superposition, this operation performed coherently will entangle the
register with the tape, just as in the third difficulty described
above.  It is no longer a problem here, since we will disentangle the
register and the tape during decoding.

Now the encoder keeps the original length tape for
{\it each\/} signal as well
as the message tape for two messages, i.e.,
$| l_1 \rangle | l_2 \rangle | h_1 h_2 0 \cdots 0\rangle_{\rm
tape} $.
Notice that it is relatively fast to compute the
length $l_1 + l_2$ of the two messages from $l_1$ and $l_2$.
Therefore, the merging procedure can be performed in polylogarithmic
depth.

More concretely, at the end the encoder obtains
\begin{equation}
| l_1 \rangle | l_2 \rangle \cdots  | l_N \rangle
| h_1 h_2 \cdots h_N 0 \cdots 0 \rangle_{\rm tape}
\end{equation}
in only $O ( (\log N)^a)$ depth for some positive integer $a$.
Finally, the encoder truncates the message tape: He keeps only say
the first
$N ( \bar{L} + \delta)$ qubits in the message tape
$| h_1 h_2 \cdots h_N 0 \cdots 0 \rangle_{\rm
tape} $ for
some $\delta > 0$ and throws away the other qubits.
This truncation minimizes the number of qubits needed.
The only overhead cost compared to the
classical case is the
storage of the length tapes of the individual signals.
This takes only $ N \lceil \log l_{\rm max} \rceil$
qubits.\footnote{Further
optimization may be possible. For instance, if $\log l_{\rm max}  $
is large, one can save
storage space by repeating
the procedure, i.e., one can now use quantum Huffman coding for
the problem of storing the quantum signals $| l_i \rangle$s.}

\subsection{Decoding}

Decoding can be done by adding an appropriate number of qubits
in the zero state $|  0 \rangle$ behind the truncated message
tape and simply running the encoding process backwards (again
with only depth $O ( (\log N)^a)$).

What about fidelity?
The key observation is
the following:

\begin{definition}
The typical subspace $S_\delta$ is the subspace where the first $N
(\bar{L} + \delta)$ qubits are arbitrary, and any qubits beyond that
are in the {\it fixed} state $| 0 \cdots 0 \rangle$.
\end{definition}

\begin{prop}
%$\forall \epsilon\ \exists \delta$ such that the fidelity $F \ge 1 -
%\epsilon$, where $F$ is the fidelity between the true state $\rho$ and
%the projection of $\rho$ on the typical subspace $S_\delta$.
%
$\forall \epsilon ,\delta > 0, \exists N_0 > 0$
such that $\forall
N > N_0$,
$F \ge 1 -
\epsilon$ where $F$ is the fidelity
between the true state
$\rho$ of the $N$ quantum signals and
the projection of $\rho$ on the typical subspace $S_\delta$
in our quantum Huffman coding scheme.

\end{prop}

Proof: The proof is identical to the case of Schumacher's noiseless
quantum coding theorem \cite{schumacher,jozsa,fuchs}.

\vspace{\baselineskip}

Therefore, the truncation and subsequent replacement of the discarded
portion by $| 0 \cdots 0 \rangle$ still lead to a high fidelity in the
decoding.

In conclusion, we have constructed an explicit parallel encoding and
decoding scheme for the storage of $N$ independent and
identically distributed quantum signals that asymptotically has
only $O ( (\log N)^a)$ depth and uses
$N ( \bar{L} + \delta + \lceil \log l_{\rm max} \rceil )$ qubits for
storage
where $ \bar{L}$ is the average length of the Huffman coding
for the classical coding problem for the set of probabilities
given by the eigenvalues of the density matrix of each signal.
Here $ \delta $ can be any positive number and $l_{\rm max}$ is the
length of the longest Huffman codeword.

\begin{cor}
A sequential implementation of the encoding algorithm requires only
$O \left(N (\log N)^a \right)$ gates.
\end{cor}

Proof: This follows immediately from the fact that the encoding is in
{\bf QNC} and uses $O(N)$ qubits: At each time step of a parallel
implementation, only $O(N)$ steps are implemented.  Since the network
has depth $O( (\log N)^a)$, there can be at most $O \left(N (\log N)^a
\right)$ gates in the network.

\section{Communication}
We now attempt to use the quantum Huffman coding for communication
rather than for the storage of quantum signals.
By communication, we assume that Alice receives the signals {\it one by
one\/}
from a source and is compelled to encode them one-by-one.
As we will show below, the
number of qubits required is slightly more, namely
$N ( \bar{L} + \delta + \lceil \log l_{\rm max} \rceil ) +
\left\lceil \log \left(N l_{\rm max}\right) \right\rceil$.
The code that we will construct is not instantaneous,
but Alice and Bob can pay a small penalty in stopping
the transmission any time.  In fact, we have the following:

\begin{theorem}
Sequential encoding and decoding of a quantum Huffman code for
communication requires only $O (N^2 (\log N)^a)$ computational
steps.
\end{theorem}

The proof follows in the next three subsections.

\subsection{Encoding}
\label{ss:sequent}
The encoding algorithm is similar to that of Section~3 except that
the signals are encoded one-by-one.
More concretely, it is done through
alternating applications of the swap and shift operations.
\begin{eqnarray}
&    & | h_1\rangle | l_1 \rangle
      | h_2\rangle | l_2 \rangle \cdots | h_N \rangle | l_N \rangle
      | {\bf 0} \rangle_{\rm
tape} \otimes \nonumber \\
& & | 0 \rangle_{\rm total~length} \nonumber \\
&\stackrel{\rm swap}{\longrightarrow} & | {\bf 0} \rangle | l_1 \rangle
      | h_2\rangle | l_2 \rangle \cdots | h_N \rangle | l_N \rangle
      | 0 \cdots 0 h_1\rangle_{\rm
tape} \otimes \nonumber \\
& & | 0 \rangle_{\rm total~length}
\nonumber \\
& \stackrel{\rm shift}{\longrightarrow} &  | {\bf 0} \rangle | l_1
\rangle
      | h_2\rangle | l_2 \rangle  \cdots | h_N \rangle | l_N \rangle
      | h_1  0 \cdots 0\rangle_{\rm
tape} \otimes \nonumber \\
& & | 0 \rangle_{\rm total~length}
\nonumber \\
& \stackrel{\rm add}{\longrightarrow} &  | {\bf 0} \rangle | l_1
\rangle
      | h_2\rangle | l_2 \rangle  \cdots | h_N \rangle | l_N \rangle
      | h_1  0 \cdots 0\rangle_{\rm
tape} \otimes \nonumber \\
& & | l_1 \rangle_{\rm total~length}
\nonumber \\
&\stackrel{\rm swap}{\longrightarrow} & | {\bf 0} \rangle | l_1 \rangle
      | {\bf 0} \rangle | l_2 \rangle   \cdots | h_N \rangle | l_N
\rangle
      | h_1  0 \cdots 0 h_2\rangle_{\rm tape} \otimes \nonumber \\
& & | l_1 \rangle_{\rm total~length} \nonumber \\
& \stackrel{\rm shift}{\longrightarrow} & | {\bf 0} \rangle | l_1
\rangle
      | {\bf 0} \rangle | l_2 \rangle   \cdots | h_N \rangle  | l_N
\rangle | h_1 h_2 0 \cdots 0 \rangle_{\rm tape} \otimes \nonumber \\
& & | l_1 \rangle_{\rm total~length} \nonumber \\
& \stackrel{\rm add}{\longrightarrow} & | {\bf 0} \rangle | l_1 \rangle
      | {\bf 0} \rangle | l_2 \rangle   \cdots | h_N \rangle  | l_N
\rangle | h_1 h_2 0 \cdots 0 \rangle_{\rm tape} \otimes \nonumber \\
& & | l_1+l_2  \rangle_{\rm total~length} \nonumber \\
&\cdots &
\nonumber \\
&\stackrel{\rm shift}{\longrightarrow} & | {\bf 0} \rangle  | l_1
\rangle
      | {\bf 0} \rangle  | l_2 \rangle  \cdots | {\bf 0} \rangle  | l_N
\rangle | h_1 h_2 \cdots h_N 0  \cdots 0 \rangle_{\rm
tape}  \otimes \nonumber \\
& & | l_1+\cdots+l_{N-1}  \rangle_{\rm total~length} \nonumber \\
&\stackrel{\rm add}{\longrightarrow} & | {\bf 0} \rangle  | l_1 \rangle
      | {\bf 0} \rangle  | l_2 \rangle  \cdots | {\bf 0} \rangle  | l_N
\rangle | h_1 h_2 \cdots h_N 0  \cdots 0 \rangle_{\rm
tape}  \otimes \nonumber \\
& & | l_1+\cdots+l_{N}  \rangle_{\rm total~length} \;.
\end{eqnarray}
We have included an ancillary space storing the total length of the
codewords generated so far.\footnote{As in equation~(\ref{merging}),
we do not include the position of the head.}  This space requires
$\log(N l_{\rm max})$ qubits.

Even though the encoding of signals themselves are done one-by-one,
the shifting operation can be sped up by parallel computation. Indeed,
as before, the required controlled-shifting operation can be performed
in $O(\log N)$ depth.  As before, if a sequential implementation is
used instead, the complete encoding of one signal still requires only
$O \left( N (\log N)^a \right)$ gates.

Now the encoding of the $N$ signals in quantum communication is done
sequentially, implying $O(N)$ applications of the shifting operation.
Therefore, with a parallel implementation of the shifting operation,
the whole process has depth $O\left( N (\log N)^a \right)$.  With a
sequential implementation, it takes $O\left( N^2 (\log N)^a \right)$
steps.

\subsection{Transmission}

Notice that the message is written on the message tape from left to
right.  Moreover, starting from left to right, the state of each qubit
once written remains unchanged throughout the encoding process. This
decoupling effect suggests that rather than waiting for the completion
of the whole encoding process, the sender, Alice, can start the
transmission immediately after the encoding.  For instance, after
encoding the first $r$ signals, Alice is absolutely sure that at least
the first $r l_{\rm min}$ (where $l_{\rm min}$ is the minimal length
of each codeword) qubits on the tape have already been written.  She
is free to send those qubits to Bob immediately. There is no penalty
for such a transmission because it is easy to see that the remaining
encoding process requires no help from Bob at all.  [Note that in the
asymptotic limit of large $r$, after encoding $r$ signals, Alice can
even send $r (\bar{L} - \epsilon)$ qubits for any $\epsilon >0$ to Bob
without worrying about fidelity.]

In addition, Alice can send the first $r$ length variables $l_1$,
$\ldots$, $l_r$, but she must retain the total-length variable for
continued encoding.  Since the total-length variable is entangled with
each branch of the encoded state, decoding cannot be completed by Bob
without use of this information.  In other words, Alice must
disentangle her system from the encoded message before decoding may be
completed.

\subsection{Decoding}
With the length information of each signal and the received qubits,
Bob can {\it start\/} the decoding process before the whole
transmission is complete {\it provided that\/} he does not perform any
measurement at this moment.  For instance, having received $r l_{\rm
min}$ qubits in the message tape from Alice, Bob is sure that at least
$s=\lfloor r l_{\rm min}/ l_{\rm max} \rfloor$ signals have already
arrived.  He can separate those $s$ signals immediately using the
length information of each signal.  This part of the decoding process
is rather straightforward and we will skip its description here.

The important observation is, however, the following:
If Bob were to perform a measurement on his signals now, he would find
that they are of poor fidelity. The reason behind this has already
been noted in Section~2. Even though the subsequent encoding process
does
not involve Bob's system, there is still entanglement between
Alice and Bob's systems. More specifically, the shifting operations in
the remaining encoding process by Alice require explicitly the
information
on the total length of decoded signals. Before Bob performs any
measurement
on his signals, it is, therefore, crucial for Alice to disentangle her
system
first, as mentioned above.

Suppose in the middle of their communication in which Bob has already
received $K \bar{L}$ qubits from Alice, Bob suddenly would like to
perform a measurement on his signals.  He shall first inform Alice of
his intention. Afterwards, one way to proceed is the following: They
choose some convenient point, say the $m$-th signal, to stop and
consider quantum Huffman coding for only the first $m$ signals and
complete the encoding and decoding processes.

We shall consider two subcases.  In the first subcase, the number $m$
is chosen such that the $m$-th signal is most likely still in the
sender (Alice)'s hands. [e.g. $m > K + O (\sqrt{K})$ in the asymptotic
limit.]  The sender Alice now disentangles the remaining signal from
the first $m$ quantum signals by applying a quantum shifting
operation.  She can now complete the encoding process for quantum
Huffman coding of the $m$ signals and send Bob any un-transmitted
qubits on the tape. In the asymptotic limit of large $K$, $O
(\sqrt{m})$ qubits of forward transmission (from Alice to Bob) are
needed.  (The required depth of the network is polynomial in $\log m$
if a parallel implementation of a quantum gate array is used.)  In
addition, Alice must send her record of the total length of the
signals.  However, this requires only an additional $\left\lceil \log
\left(m l_{\rm max}\right) \right\rceil$ qubits, so the total number
which must be transmitted for disentanglement is still $O (\sqrt{m})$.

In the second subcase, the number $m$ is chosen such that the $m$-th
signal is most likely already in the receiver (Bob)'s hands.  [e.g. $m
< K - O (\sqrt{K})$ in the asymptotic limit.]  The receiver Bob now
attempts to disentangle the remaining signals from the first $m$
quantum signals by applying a quantum shifting operation.  Of course,
he needs to shift some of his qubits back to Alice.  This
asymptotically amounts to $O (\sqrt{m})$ qubits of {\it backward\/}
communication.  This is a penalty that one must pay for this method.
After this is done, Alice must again send her length register to Bob
(after subtracting the lengths of the signals returned to her).  This
requires an additional $O(\log m)$ qubits.

If $m$ is chosen between $K - O(\sqrt{K})$ and $K + O(\sqrt{K})$,
neither sending signals forward or backward will suffice to
properly disentangle the varying lengths of the signals.  One possible
solution is to choose $m' > K + O(\sqrt{K})$ and perform the above
procedure, sending $m'$ total signals to Bob.  Then Bob decodes and
returns the $m' - m$ extra signals to Alice.  This method requires
$O(\sqrt{K})$ qubits transmitted forward and $O(\sqrt{K})$ qubits
transmitted backwards to disentangle.

We remark that the shifting operation can be done rather easily in
distributed quantum computation between Alice and Bob. This is a
non-trivial observation because the number of qubits to be shifted
from Alice to Bob is itself a quantum mechanical variable. This,
however, does not create much problem.  Bob can always communicate
with Alice using a bus of fixed length.  For example, he applies local
operations to swap the desired quantum superposition of various
numbers of qubits from his tape to the bus, sends such a bus to Alice,
etc.

The result is the following theorem:

\begin{theorem}
Alice and Bob may truncate a communication session after the
transmission of $m$ encoded signals, retaining high fidelity with the
cost of $O(\sqrt{m})$ additional qubits transmitted.
\end{theorem}

In the above discussion, we have focused on the simple case when Bob
would like to perform a measurement on
the whole set of the first $m$ signals. Suppose Bob is interested only
in a particular signal, say the $m$-th one, but not the others. There
exists a more efficient scheme for doing it. We shall skip
the discussion here.

\section{Concluding Remarks}
We have successfully constructed a Huffman-coding inspired scheme
for the storage of quantum information. Our scheme is
highly efficient. The encoding and
decoding processes of $N$ quantum signals can be done {\it in
parallel\/}
with depth polynomial in
$\log N$. (If parallel machines are unavailable,
as shown in subsection~\ref{ss:sequent}
our encoding scheme will
still take only $O \left(N (\log N)^a \right)$ computational steps for a
sequential implementation.
In contrast, a naive
implementation of Schumacher's scheme will
require $O(N^3)$ computational steps.)
This massive parallelism is possible because we explicitly use another
tape to store the length information of the individual signals.
The storage space needed is asymptotically
$N ( \bar{L} + \delta + \lceil \log l_{\rm max} \rceil )$
where $\bar{L}$ is the average length of the corresponding classical
Huffman coding problem for the density matrix in the
diagonal form,
$\delta $ is an arbitrary small positive
number and $ l_{\rm max}$ is the length of the longest Huffman codeword.

We also considered the problem of using quantum Huffman coding for
communication in which case Alice encodes the signals one-by-one.  $N
( \bar{L} + \delta +\lceil \log l_{\rm max} \rceil ) + O(\log N)$
qubits are needed.  With a parallel implementation of the shifting
operation, depth of $O (N (\log N)^a)$ is needed.  On the other hand,
with a sequential implementation, $O (N^2 (\log N)^a)$ computational
steps are needed.  In either case, the code is not instantaneous, but,
by paying a small penalty in terms of communication and computational
costs, Alice and Bob have the option of stopping the transmission and
Bob may then start measuring his signals.

More specifically, while the receiver Bob is free to separate the
signals from one another, he is not allowed to measure them until the
sender Alice has completed the encoding process.  This is because
Alice's encoder head generally contains the information of the total
length of the signals. In other words, its state is entangled with
Bob's signals. Therefore, whenever Bob would like to perform a
measurement, he should first inform Alice and the two should proceed
with disentanglement. We present two alternative methods of achieving
such disentanglement one of which involves forward communication and
the other of which involves both forward and backward.

Since real communication channels are always noisy,
in actual implementation source coding is always followed by encoding
into an error correcting code. Following the pioneering work
by Shor \cite{shor} and independently by Steane \cite{steane},
various quantum error correcting codes have been constructed.
We remark that quantum Huffman coding algorithm (even the version for
communication) can be
immediately combined with the encoding process
of a quantum error correcting code for efficient
communication through a noisy channel.

As quantum information is fragile against noises in the environment,
it may be useful to work out a fault-tolerant procedure for quantum
source coding. The generalizations of other classical coding schemes
to the quantum case are also interesting \cite{jhhh}. Moreover, there
exist universal quantum data compression schemes motivated by the
Lempel-Ziv compression algorithm for classical
information~\cite{private}.

\section{Acknowledgment}
One of us (H.-K. Lo) thanks D. P. DiVincenzo, J. Preskill and T.
Spiller for helpful discussions. This work is supported in part by
EPSRC grants GR/L91344 and GR/L80676, by the Lee A. DuBridge
Fellowship, by DARPA under Grant No. DAAH04-96-1-0386 through the
Quantum Information and Computing (QUIC) Institute administered by
ARO, and by U.S. Department of Energy under Grant No.
DE-FG03-92-ER40701.

\end{document}